\documentclass{moriond}

\def\be{\begin{equation}}
\def\ee{\end{equation}}
\def\bea{\begin{eqnarray}}
\def\eea{\end{eqnarray}}

\newcommand{\Photo}{}

\begin{document}
\vspace*{4cm}
\title{Towards N$^3$LO PDFs and their implications for Higgs production cross-sections}

\author{ Felix Hekhorn }

\address{University of Jyvaskyla, Department of Physics, P.O. Box 35, FI-40014 University of Jyvaskyla, Finland\\
Helsinki Institute of Physics, P.O. Box 64, FI-00014 University of Helsinki, Finland}

\maketitle\abstracts{
We review the recently published PDF sets \texttt{MSHT20xNNPDF40\_an3lo} and \texttt{MSHT20xNNPDF40\_an3lo\_qed}, which have been obtained
by combining the approximate N$^3$LO (aN$^3$LO) sets of parton distributions (PDFs) from the MSHT20 and NNPDF4.0 series.
We investigate the impact of the most recent results on N$^3$LO splitting functions onto the two PDF sets.
Finally, we demonstrate the impact of N$^3$LO PDFs on the computation of Higgs production cross-sections at the LHC.
}

\section{Fitting PDFs at aN$^3$LO}

In Ref.~\cite{MSHT:2024tdn} two new PDF sets, \texttt{MSHT20xNNPDF40\_an3lo} and \texttt{MSHT20xNNPDF40\_an3lo\_qed}, are presented which have been obtained
by combining the approximate N$^3$LO (aN$^3$LO) sets of parton distributions (PDFs) from the MSHT20~\cite{Bailey:2020ooq} and
NNPDF4.0~\cite{NNPDF:2021njg} series.
Specifically, we used either the pure QCD extractions~\cite{McGowan:2022nag,NNPDF:2024nan}
or the variants including in addition QED corrections~\cite{Cridge:2023ryv,Barontini:2024dyb}.
The combination is performed by drawing 100 replicas respectively from the publicly available PDF sets and summing them
plainly together, such that the new set is a distribution covering both extractions.

When considering the perturbative order in a PDF fit it is instructive to recall the general structure on how to compute
theory predictions.
For an observable $F$, which depends linearly on the fitted PDF $\mathbf f(Q_0)$,
e.g.\ structure functions in deep-inelastic scattering (DIS), we find
\begin{equation} \label{eq:fact}
    F(Q^2) = \mathbf C(Q^2) \mathbf E^{(5)}(Q^2 \leftarrow m_b^2) \mathbf A^{(4)}(m_b^2) \mathbf E^{(4)}(m_b^2 \leftarrow Q_0^2) \mathbf f^{(4)}(Q_0^2)
\end{equation}
where $\mathbf E$ denotes the evolution kernel operator (EKO), i.e.\ the solution of the DGLAP equations, $\mathbf A$ denotes
the massive operator matrix elements\footnote{a.k.a.\ transition matrix elements} (OMEs), which transform the PDF here between between the four
and five flavour scheme, and $\mathbf C$ denotes the partonic matrix elements.
While ideally one would consider all these elements at N$^3$LO accuracy to obtain N$^3$LO accurate PDFs, this is in practice
not possible as not all ingredients are known at that accuracy and each PDF group had to define a suitable
approximation strategy - we refer the reader to the respective publications for a detailed discussion.
While there has been relevant progress in recent years on computing splitting functions, which are needed to
compute the EKOs $\mathbf E$, and OMEs $\mathbf A$, the computation of partonic matrix elements $\mathbf C$ remains a long term project
which poses many challenges both on conceptual and practical matters~\cite{Caola:2022ayt}.
In fact, DIS structure functions, i.e.\ the example of Eq.~(\ref{eq:fact}), are one of the few cases in which the
(massless) coefficient functions $\mathbf C$ are exactly known at present~\cite{Vermaseren:2005qc}.
We recall that the starting scale of MSHT20 fits is in a 3 flavor regime, thus in comparison to Eq.~(\ref{eq:fact}) which holds for NNPDF,
an additional EKO $\mathbf E^{(3)}$ and a additional OMEs $\mathbf A^{(3)}$ are needed for computing observables.

\section{Preliminary results using most recent splitting functions}

Since the release of the PDF sets, which are underlying the combination, significant efforts have been reached in obtaining more perturbative
ingredients.
Here, we investigate on how these new results impact the PDF extraction of the two groups.
In particular, Ref.~\cite{Falcioni:2024qpd} (FHMRUVV) obtained the currently best available parametrization of the aN$^3$LO splitting functions
and Ref.~\cite{Ablinger:2024xtt} the currently best available OMEs.
In addition, a variant of the MSHT20 aN$^3$LO fit has been produced which now uses the same OMEs as in NNPDF4.0.
In this way, the updated MSHT20 and NNPDF4.0 aN$^3$LO fits use the same ``known'' perturbative information on N$^3$LO calculations, and in terms of theory
input they only differ in the treatment of ``unknown''  N$^3$LO effects such as the massive matrix elements for the DIS structure functions or the N$^3$LO $K$-factors for hadronic processes.

\begin{figure}[h]
  \centering
  \includegraphics[width=0.48\textwidth]{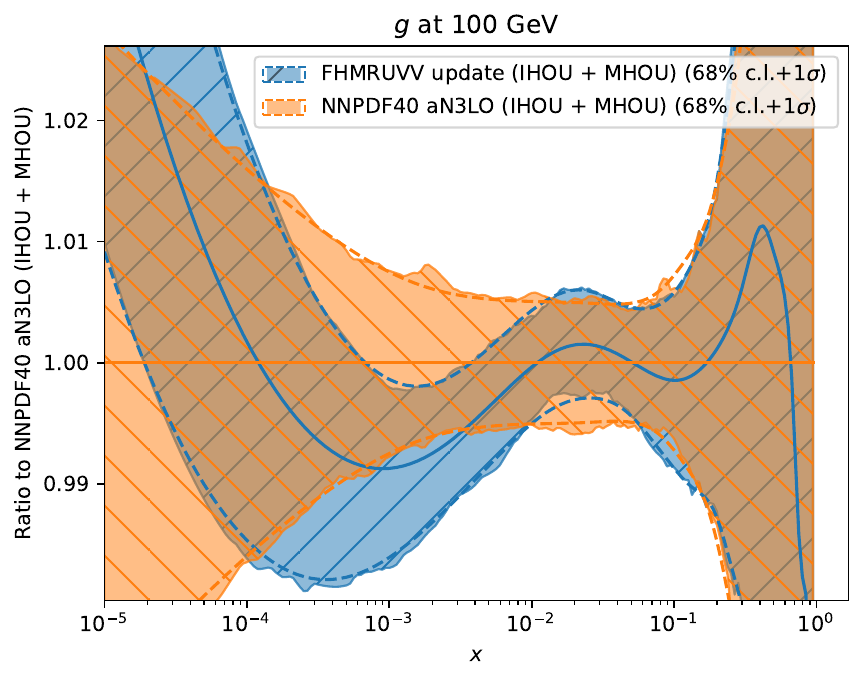}%
  \includegraphics[width=0.48\textwidth]{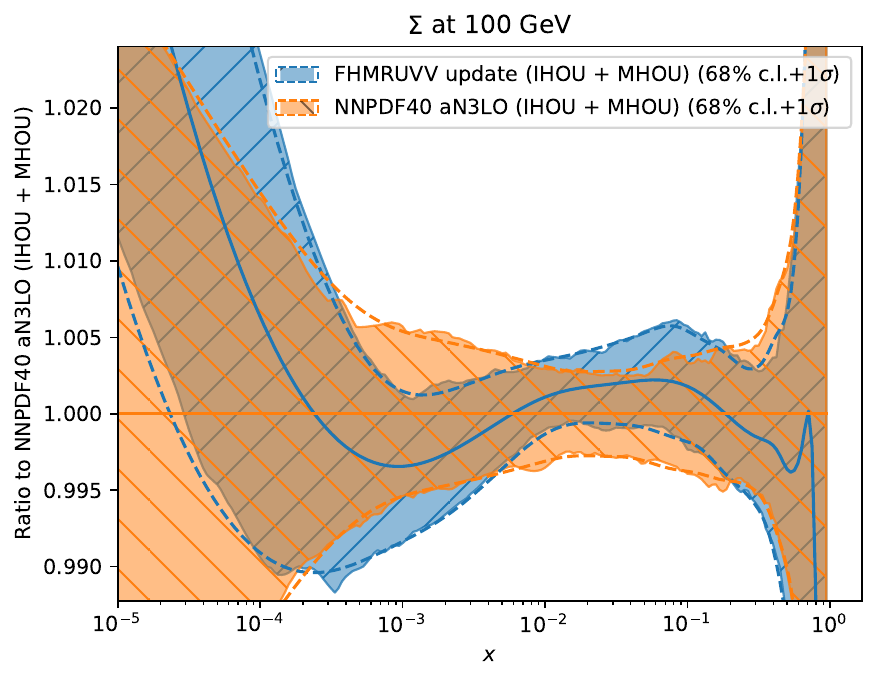}
  \caption{The impact of using the latest FHMRUVV parametrisation of the N$^3$LO  splitting functions
            and the lastest OMEs in the NNPDF4.0 aN$^3$LO fit, for
            the gluon (left) and total quark singlet (right) PDFs at $Q=100$ GeV. }
  \label{fig:nnpdf40_updated}
\end{figure}

To investigate the impact of using the latest FHMRUVV parametrisation of the N$^3$LO  splitting functions and of
the N$^3$LO heavy quark OMEs a new MSHT20 aN$^3$LO fit has been performed~\cite{MSHTatDIS}.
Similarly a new NNPDF4.0 aN$^3$LO fit has been obtained:
Fig.~\ref{fig:nnpdf40_updated} displays the impact of using the latest FHMRUVV parametrisation of the N$^3$LO  splitting functions in the
NNPDF4.0 aN$^3$LO fit, for the gluon (left) and total quark singlet (right) PDFs at $Q=100$ GeV.
For the both PDFs we find that the differences in the new central value compared to the baseline results are always contained within the PDF uncertainties.

We conclude that adopting a common set of FHMRUVV splitting functions and also heavy quark OMEs leads to quite moderate
effects at the PDF level and hence it does not change the conclusions of the original combination concerning predictions for Higgs cross-sections at the LHC. 
In order to study the impact of the updated PDF sets, we have also produced a new combination based on these variants, which we consider in the next section.

\section{Impact on Higgs cross-sections}

\begin{figure}[h]
    \centering
    \includegraphics[width=0.49\textwidth]{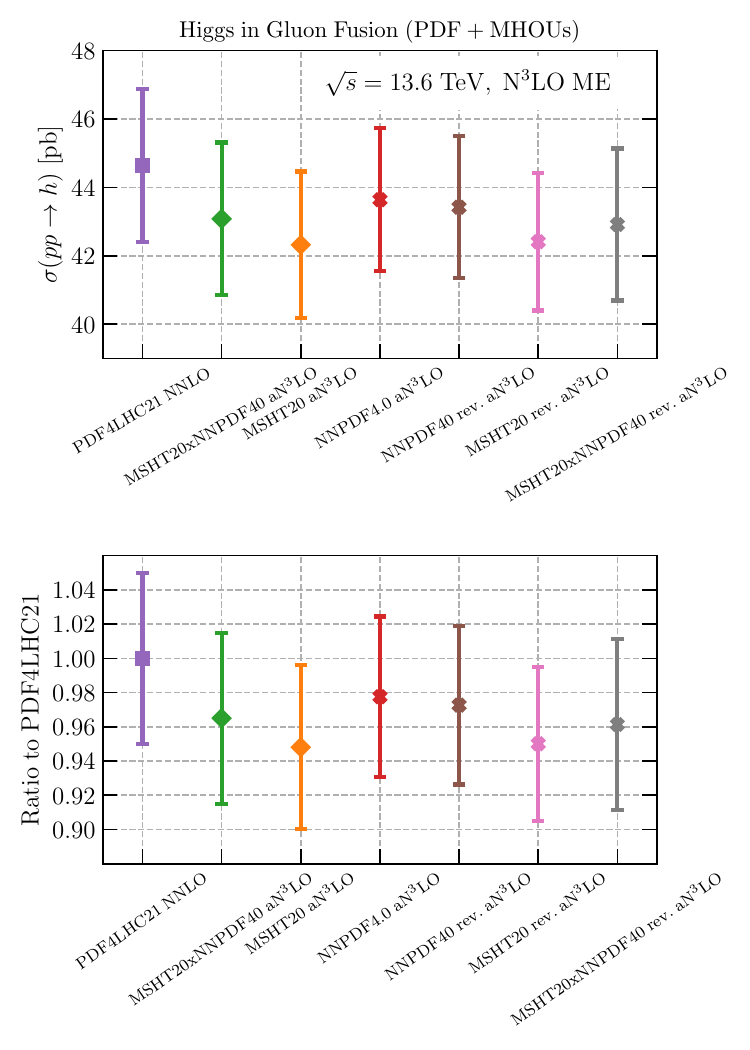}
    \includegraphics[width=0.49\textwidth]{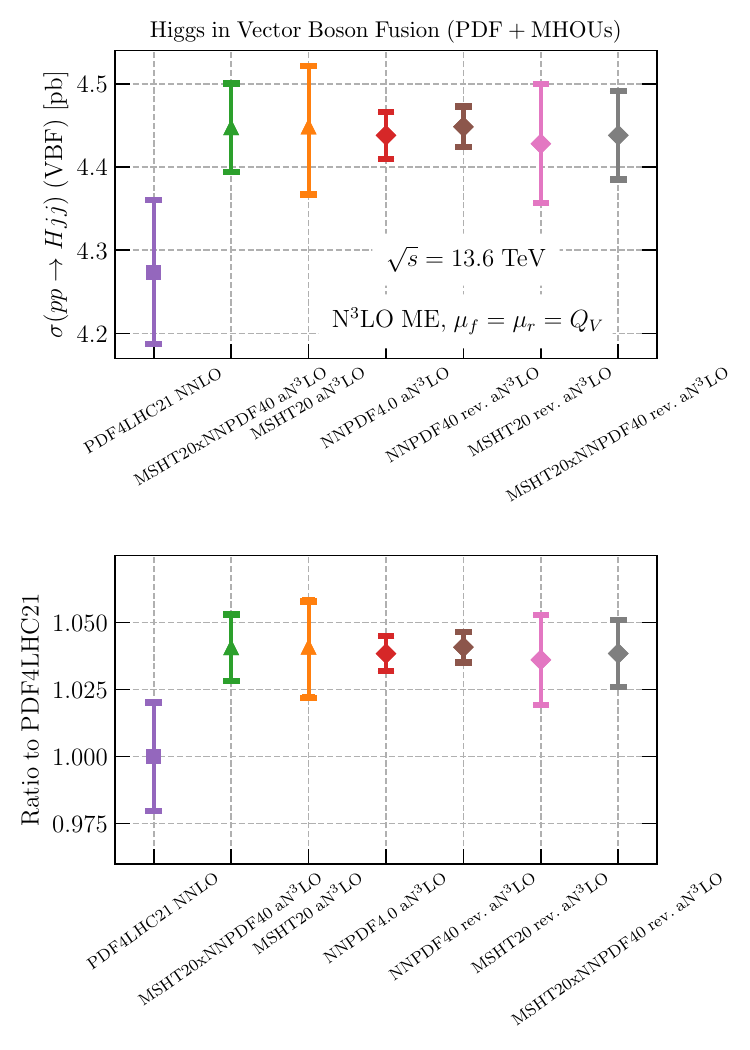}\\
    \caption{The Higgs production cross-section in gluon fusion (left panels) and in vector-boson fusion (right panels) at $\sqrt{s}=13.6$ TeV, where
    we compare the original aN$^3$LO PDF sets and their combination, with the revised (``rev.'') version discussed here and the associated combination. }
    \label{fig:higgs-ggF-n3lo-HXSWG.pdf}
\end{figure}

In Fig.~\ref{fig:higgs-ggF-n3lo-HXSWG.pdf} we show two relevant Higgs production cross-sections as e.g.\ measured at the LHC.
We compare predictions between the PDF sets from PDF4LHC21~\cite{PDF4LHCWorkingGroup:2022cjn}, the original combination~\cite{MSHT:2024tdn} and their
underlying PDF sets, and the preliminary results presented here.
Comparing to Ref.~\cite{MSHT:2024tdn}, the predictions for Higgs production via gluon-gluon fusion move down by about $0.5\%$ for NNPDF and up
by slightly under $0.5\%$ for MSHT (the change was about $0.7\%$ when only the splitting functions were updated).
For Higgs production via vector boson fusion the changes are of order $0.1\%$ for both groups.
Hence, for gluon-gluon fusion the use of common splitting functions and OMEs has a tiny effect on the best prediction of the Higgs cross section
compared to the combination using the published PDF sets.
Indeed, the fact that the PDFs move slightly together and that the theoretical uncertainties will automatically be improved by the reduction in 
uncertainty in N$^3$LO splitting functions and (in the case of MSHT20) OMEs suggests that the uncertainty in an updated combination would decrease,
and hence, the uncertainty in our official combination may be taken as conservative.

We conclude that the combination presented in Ref.~\cite{MSHT:2024tdn} is reliable and can be used to study the
effect of N$^3$LO PDFs at the LHC or the EIC.
The preliminary PDFs and results discussed here only present a first phenomenological study and we defer a detailed
investigation to future publications.

\section*{Acknowledgments}

F.~H. is supported by the Academy of Finland project 358090 and is funded as a
part of the Center of Excellence in Quark Matter of the Academy of Finland, project 346326.

\section*{References}
\bibliography{moriond}

\providecommand{\href}[2]{#2}\begingroup\raggedright\begin{thebibliography}{10}

\bibitem{MSHT:2024tdn}
{\bf MSHT, NNPDF} Collaboration, T.~Cridge et~al., {\it {Combination of aN$^3$LO PDFs and implications for Higgs production cross-sections at the LHC}},  \href{http://arxiv.org/abs/2411.05373}{{\tt arXiv:2411.05373}}.

\bibitem{Bailey:2020ooq}
S.~Bailey, T.~Cridge, L.~A. Harland-Lang, A.~D. Martin, and R.~S. Thorne, {\it {Parton distributions from LHC, HERA, Tevatron and fixed target data: MSHT20 PDFs}},  {\em Eur. Phys. J. C} {\bf 81} (2021), no.~4 341, [\href{http://arxiv.org/abs/2012.04684}{{\tt arXiv:2012.04684}}].

\bibitem{NNPDF:2021njg}
{\bf NNPDF} Collaboration, R.~D. Ball et~al., {\it {The path to proton structure at 1\% accuracy}},  {\em Eur. Phys. J. C} {\bf 82} (2022), no.~5 428, [\href{http://arxiv.org/abs/2109.02653}{{\tt arXiv:2109.02653}}].

\bibitem{McGowan:2022nag}
J.~McGowan, T.~Cridge, L.~A. Harland-Lang, and R.~S. Thorne, {\it {Approximate N$^{3}$LO parton distribution functions with theoretical uncertainties: MSHT20aN$^3$LO PDFs}},  {\em Eur. Phys. J. C} {\bf 83} (2023), no.~3 185, [\href{http://arxiv.org/abs/2207.04739}{{\tt arXiv:2207.04739}}]. [Erratum: Eur.Phys.J.C 83, 302 (2023)].

\bibitem{NNPDF:2024nan}
{\bf NNPDF} Collaboration, R.~D. Ball et~al., {\it {The path to $\hbox {N}^3\hbox {LO}$ parton distributions}},  {\em Eur. Phys. J. C} {\bf 84} (2024), no.~7 659, [\href{http://arxiv.org/abs/2402.18635}{{\tt arXiv:2402.18635}}].

\bibitem{Cridge:2023ryv}
T.~Cridge, L.~A. Harland-Lang, and R.~S. Thorne, {\it {Combining QED and approximate ${\rm N}^3$LO QCD corrections in a global PDF fit: MSHT20qed\_an3lo PDFs}},  {\em SciPost Phys.} {\bf 17} (2024), no.~1 026, [\href{http://arxiv.org/abs/2312.07665}{{\tt arXiv:2312.07665}}].

\bibitem{Barontini:2024dyb}
A.~Barontini, N.~Laurenti, and J.~Rojo, {\it {NNPDF4.0 aN$^3$LO PDFs with QED corrections}},  in {\em {31st International Workshop on Deep-Inelastic Scattering and Related Subjects}}, 6, 2024.
\newblock \href{http://arxiv.org/abs/2406.01779}{{\tt arXiv:2406.01779}}.

\bibitem{Caola:2022ayt}
F.~Caola, W.~Chen, C.~Duhr, X.~Liu, B.~Mistlberger, F.~Petriello, G.~Vita, and S.~Weinzierl, {\it {The Path forward to N$^3$LO}},  in {\em {Snowmass 2021}}, 3, 2022.
\newblock \href{http://arxiv.org/abs/2203.06730}{{\tt arXiv:2203.06730}}.

\bibitem{Vermaseren:2005qc}
J.~A.~M. Vermaseren, A.~Vogt, and S.~Moch, {\it {The Third-order QCD corrections to deep-inelastic scattering by photon exchange}},  {\em Nucl. Phys. B} {\bf 724} (2005) 3--182, [\href{http://arxiv.org/abs/hep-ph/0504242}{{\tt hep-ph/0504242}}].

\bibitem{Falcioni:2024qpd}
G.~Falcioni, F.~Herzog, S.~Moch, A.~Pelloni, and A.~Vogt, {\it {Four-loop splitting functions in QCD \textendash{} the gluon-gluon case \textendash{}}},  {\em Phys. Lett. B} {\bf 860} (2025) 139194, [\href{http://arxiv.org/abs/2410.08089}{{\tt arXiv:2410.08089}}].

\bibitem{Ablinger:2024xtt}
J.~Ablinger, A.~Behring, J.~Bl\"umlein, A.~De~Freitas, A.~von Manteuffel, C.~Schneider, and K.~Sch\"onwald, {\it {The non-first-order-factorizable contributions to the three-loop single-mass operator matrix elements AQg(3) and \ensuremath{\Delta}AQg(3)}},  {\em Phys. Lett. B} {\bf 854} (2024) 138713, [\href{http://arxiv.org/abs/2403.00513}{{\tt arXiv:2403.00513}}].

\bibitem{MSHTatDIS}
T.~Cridge et~al., ``{Approximate N3LO PDFs: Updates and Consequences for Phenomenology}.'' To appear in DIS2025, 2025.

\bibitem{PDF4LHCWorkingGroup:2022cjn}
{\bf PDF4LHC Working Group} Collaboration, R.~D. Ball et~al., {\it {The PDF4LHC21 combination of global PDF fits for the LHC Run III}},  {\em J. Phys. G} {\bf 49} (2022), no.~8 080501, [\href{http://arxiv.org/abs/2203.05506}{{\tt arXiv:2203.05506}}].

\end{thebibliography}\endgroup
\end{document}